\documentclass[twocolumn,showpacs,preprintnumbers,amsmath,amssymb,superscriptaddress]{revtex4-1}
\usepackage[english]{babel}
\usepackage{mathrsfs}
\usepackage{epsfig}
\usepackage{multirow}
\usepackage{graphicx,array}
\def\p{\partial}
\def\bx{\bgroup \bf x\egroup}
\def\bn{\textbf{n}}

\def\bt{\textbf{t}}
\def\bp{\textbf{p}}
\def\const{\rm const}
\def\ve{\varepsilon}

\def\tr{\mathop{\rm tr}\nolimits}
\def\am{\mathop{\rm am}\nolimits}
\def\Ai{\mathop{\rm Ai}\nolimits}
\def\be{\begin{equation}}
\def\ee{\end{equation}}

\begin{document}

\title{Protein crowding on biomembranes: analysis of contour instabilities}

\author{O. V. Manyuhina} 
\affiliation{Nordita, The Royal Institute of Technology \& Stockholm University,  Roslagstullsbacken~23, SE-10691 Stockholm, Sweden} 
\affiliation{Physics Department, Syracuse University, Syracuse, NY 13244, USA}
\date{\today}

\begin{abstract}
Collective behavior of proteins on biomembranes is usually studied within the spontaneous curvature model. Here we consider an alternative phenomenological approach, which accounts consistently for partial  ordering of proteins as well as the anchoring forces exerted on a membrane by  layer of proteins. We show analytically that such anisotropic interactions can drive membrane bending, resulting in non-trivial equilibrium morphologies. The predicted instabilities can advance our conceptual understanding of  physical mechanisms behind collective phenomena  in biological systems, in particular those with inherent anisotropy.
\end{abstract}

\maketitle

%
\section{Introduction}

Composed of lipids and proteins, membranes of cells and organelles are  known to play an important role in many biological processes, and are integral to endocytosis, cytokinesis, apoptosis, fusion and fission of cells~\cite{mcmahon:2005,kozlov:2006}. On the one hand, these complex processes are accompanied by bending of biomembranes, which may result in the budding of vesicles (endocytosis). On the other hand, the observed shape transformations of membranes are intertwined with the collective behavior of proteins. It is believed that the ability of proteins to generate and to sense curvature(s)~\cite{mynote1}  of biomembrane are crucial in orchestrating budding of vesicles~\cite{mcmahon:2005,kozlov:2006,sorre:2012}. In particular, two main mechanisms of proteins binding to a membrane, which can generate curvature are due to i)~membrane scaffolding by intrinsically curved proteins or ii)~insertion of wedge-like amphiphatic helix into bilayer~\cite{mcmahon:2005,kozlov:2006}. Recent experiments~\cite{stachowiak:2012}, however, have tested the hypothesis of clathrin-mediated endocytosis and suggested that protein--protein crowding may be a third  mechanism for bending the membrane.  

Most theoretical models, explore budding phenomena within the Canham--Helfrich spontaneous curvature model~\cite{helfrich:1973}. Such simplified approach assumes coupling between the local density of proteins and the mean curvature of a lipid bilayer, resulting in spatially inhomogeneous spontaneous curvature, which either depends linearly on the density of bound proteins~\cite{leibler:1986,benamar:2004}, or corresponds to a hypothetical distribution of proteins~\cite{agrawal:2009}, or three different spontaneous curvatures are prescribed to model the effective interactions of anisotropic proteins with membrane~\cite{akabori:2011}. The  phase-field approach considered in~\cite{jordi:2013} explores the role of protein--lipid affinity and non-equilibrium dynamics on the clustering of proteins. In turn, the statistical mechanics approach proposed in~\cite{martin:2012} deals with the entropic effects due to the crowding of proteins, modeled as liquid composed of hard disks. Although proteins may be considered in liquid state inside the cell, on the surface of the liquid--crystalline bilayer they become partly ordered. The proposed approaches are perhaps too simple to discuss consistently over different lengthscales the influence of membrane morphologies on the partial ordering of proteins and {\it vice versa}.

In this paper we consider an alternative phenomenological model where proteins are i)~envisaged as a continuum layer attached to one side of a biomembrane, ii)~described by a local orientational order and density per unit area. Interactions between membrane and proteins are modeled as anisotropic surface tension, dependent on the surface normal,  also known as anchoring in theory of liquid crystals~\cite{lavrent:book}. The membrane mediated interactions of two different inclusions were studied before~\cite{biscari:2002} in the limit of strong anchoring (fixed angle between the normal to the membrane and the inclusion axis at the contact point), leading to the change of an optimal shape of a membrane. An exact solution to the global non-linear bending phenomena of a spherical vesicle in presence of a grafted latex bead, was derived in~\cite{benoit:2007}. Here instead we consider a `crowded' state i.e. membrane is coated by proteins, which interact with each other either through excluded volume interactions (Onsager model of hard rods) or anisotropic van der Waals forces of attraction (Maier and Saupe theory), both resulting in the emergence of nematic-like phase with long-range orientational order. Contrary to widely studied nematic shells with in-plane orientational order~\cite{biscari:2006,bowick:2009,napoli:2012}, here we do not confine proteins to the tangent plane of the surface, instead we allow them to approach biomembrane at some angle  relative to the normal, which is not fixed {\it a priori}. Because of the competing anchoring and bending energy effects, the projection of protein length on the tangent plane varies along the surface. \looseness=-1

In the following, we formulate a one-dimensional version of the model and explore analytically the equilibrium shapes of biomembrane.  Without referring to any specific experiments we illustrate the role of anchoring forces on the bending of biomembranes. Our analysis suggests that the proposed minimal model is a viable approach for studying the mechanics of biomembranes in presence of proteins.

\section{Theoretical formalism}

\begin{figure}[bt]
\centering
\includegraphics[width=0.7\linewidth]{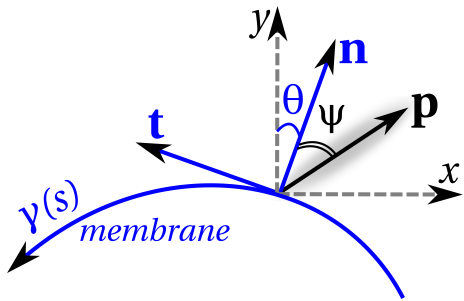}
\caption{Membrane is described by the curve $\gamma(s)$, where~$s$ is the curvilinear coordinate. Two systems of coordinates are introduced: the global Cartesian one along $x$- and $y$-axis, and the local one, formed by the unit normal $\bn$ and the unit tangent vector $\bt$ to the membrane. The parametrization is chosen such as $\bt=\gamma'/|\gamma'| = (-\cos\theta,\sin\theta)$. The average orientation of attached proteins is described by the vector  $\bp= \cos\psi\,\bn - \sin\psi \,\bt$. The normal $\bn$ points out in the direction of proteins $\bp$, approaching the membrane either from inside or outside of the cell.}
\label{fig:protein}
\end{figure}

The membrane is modeled as a curve: $s\to \gamma(s)=(x(s),y(s))$ in the $xy$-plane parametrized by the local angle $\theta(s)$ between the normal $\bn$ and the $y$-axis (see~Fig.~\ref{fig:protein}). The curvature of the membrane is $|\gamma''(s)|=\theta'$, where prime  denotes the derivative with respect to $s$, yielding the bending energy  per unit length as ${\cal E}_{\rm bend}=\kappa/2\int_\gamma ds\,(\p_s \theta)^2$, where $\kappa$ is the bending rigidity of the membrane, $\kappa\propto1-10~k_B T$~\cite{helfrich:1973}.

Proteins are modeled as anisotropic elongated molecules attached to one side of biomembrane (see Fig.~\ref{fig:protein}). For protein ensembles with an averaged orientation $\bp$, the relevant order parameter is the traceless 2-rank tensor ${\bf Q} = q \big(\bp\otimes \bp-\frac{\bf I}2\big)$. Both $\bp$ and $q$ are functions of the arclength $s$ associated with the membrane. Field $q$ measures the degree of orientational order in the theory of nematic--isotropic phase transition~\cite{lavrent:book}. Within our framework, $q$ can be also conceived as weighted mass distribution of proteins on a membrane. Similar to the lyotropic liquid crystals~\cite{lavrent:book}, we assume that there is a critical density of proteins $\rho^*$ below which $q=0$  (protein `free') and above which we have $q>0$ (protein `crowded'). In the vicinity of the transition ($\rho=\rho^*$),  we assume the  Landau-deGennes form~\cite{lavrent:book,mottram:2004} for the free energy of proteins bound to one side of membrane, written as
\be\label{eq:ldg}
{\cal E}_{\rm LdG}=\int_\gamma ds\, \big\{\kappa_p |\p_s {\bf Q}|^2 + a \tr({\bf Q}^2)+ c \tr({\bf Q}^4)\big\},
\ee
where only the coefficient $a\propto(\rho^*-\rho)$ depends on the protein density $\rho$, while $c>0$ is assumed to be constant. In the local system of coordinate the averaged orientation of proteins is $\bp= \cos\psi\,\bn - \sin\psi \,\bt$. Since  $\bt' =\theta'\bn$ and $\bn'=-\theta'\bt$, it follows that $|\p_s\bp|^2=(\theta'+\psi')^2$, $|\p_s{\bf Q}|^2=q'^2/2+2q^2(\theta'+\psi')^2$, and $\tr ({\bf Q}^2)=q^2/2$,  $\tr ({\bf Q}^4)=q^4/8$. Based on the van der Waals attraction between two protein rods in solution and  value of the Hamaker constant for proteins in water~\cite{israel:book,maha:kinks}, one can estimate the elastic constant $\kappa_p\propto1-10~k_BT$ entering~\eqref{eq:ldg}, which has the same order of magnitude as the bending rigidity~$\kappa$.

Finally, we assume that proteins favor  to approach the membrane at some specific orientation $\bp_0$ (`easy axis') or angle $\psi_0$ relative to the normal $\bn$. Then the leading order contribution to the  interaction between proteins and the biomembrane is proportional to $\bp_0\cdot {\bf Q}\cdot\bp_0^T$, which can be cast into the form ${\cal E}_{\rm anch}=\omega/2\int_\gamma ds\, q \sin^2(\psi(s)-\psi_0)^2$, known as anchoring~\cite{lavrent:book,mottram:2004}. For example, network of the cortical actin filaments approaches the lipid membrane almost tangentially~\cite{benamar:2011}, so that $\psi_0\to \pm\pi/2$.  
The anchoring strength $\omega$ for proteins on biomembrane may vary within a wide range depending on the local density of proteins and type of chemical bonding; for nematic liquid crystals $\omega\simeq 10^{-7}-10^{-4}~$J/m$^2$. Next we explore the consequences of this simplified model and illustrate how the crowding effect of partly ordered proteins can drive bending of biomembranes.

\section{Anchoring effects}

\hskip-1ex\begin{table*}[htb]
\caption{\label{tab:shapes}Equilibrium shapes of biomembrane $\gamma(s)$ in presence of attached proteins, calculated by numerical integration of $x(s)=-\int_0^s d\sigma\,\cos\theta(\sigma)$ and $y(s)=\int_0^s d\sigma\,\sin\theta(\sigma)$ using~\eqref{eq:sol1}, we choose $\tilde\kappa=1$, $n_{\theta}=1$. The dimensionless anchoring strength  $\tilde\omega=R\sqrt{ 2\omega \bar q/\kappa}$ is a sole parameter responsible for the instability towards elongated shapes ($n_\psi=1$) or periodically undulated shapes ($n_\psi=2,3$), which occurs at $\tilde\omega\simeq10$. The color of the closed curves is associated with the protein tilt relative to $\psi_0$, scaled to $\sin(\psi-\psi_0)$ on the colorbar.} 
{\def\tabcolsep{16pt}
\centering
\begin{tabular}{m{1mm}m{33mm}m{33mm}m{33mm}m{0.1ex}}
& \centering $\tilde\omega=1$&\centering  $\tilde\omega=10$& \centering $\tilde\omega=30$ &\\[3ex] 
${n_\psi=1}$& & & &\\[-4ex]
&\centering \includegraphics[height=3.cm]{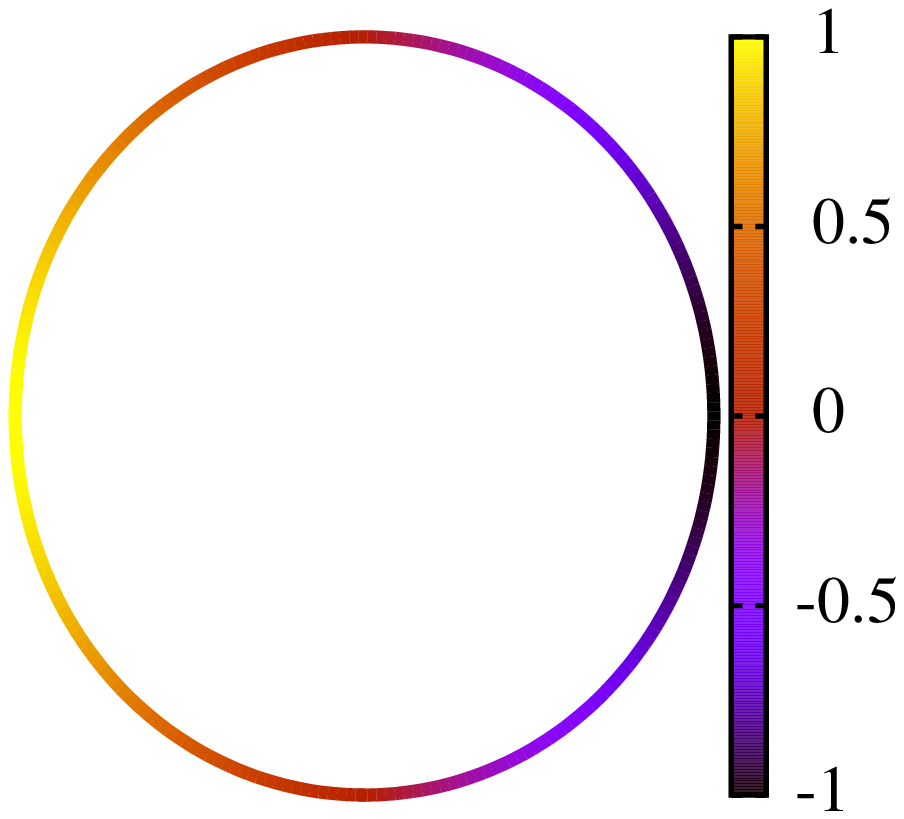}
&\centering \centering\includegraphics[height=3.cm]{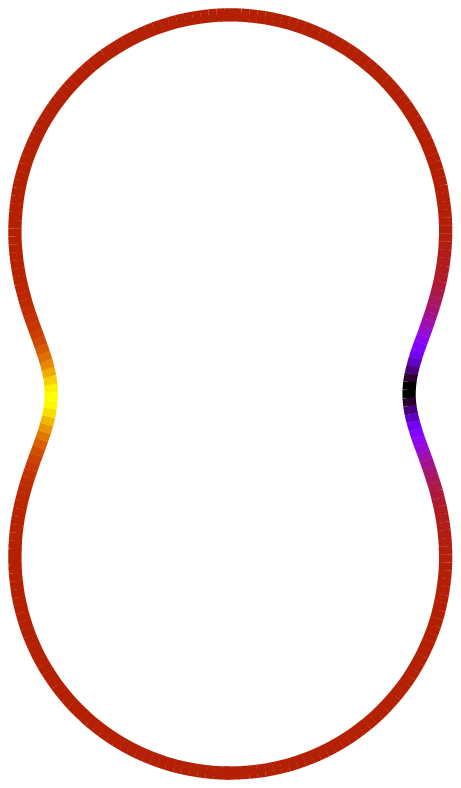}
&\centering \includegraphics[height=3.cm]{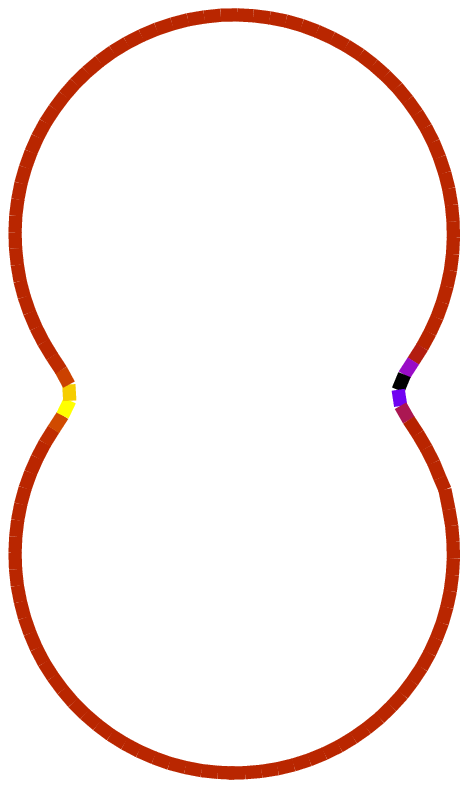}
&\\
${n_\psi=2}$& & & &\\[-4ex]
&\centering \includegraphics[height=3.cm]{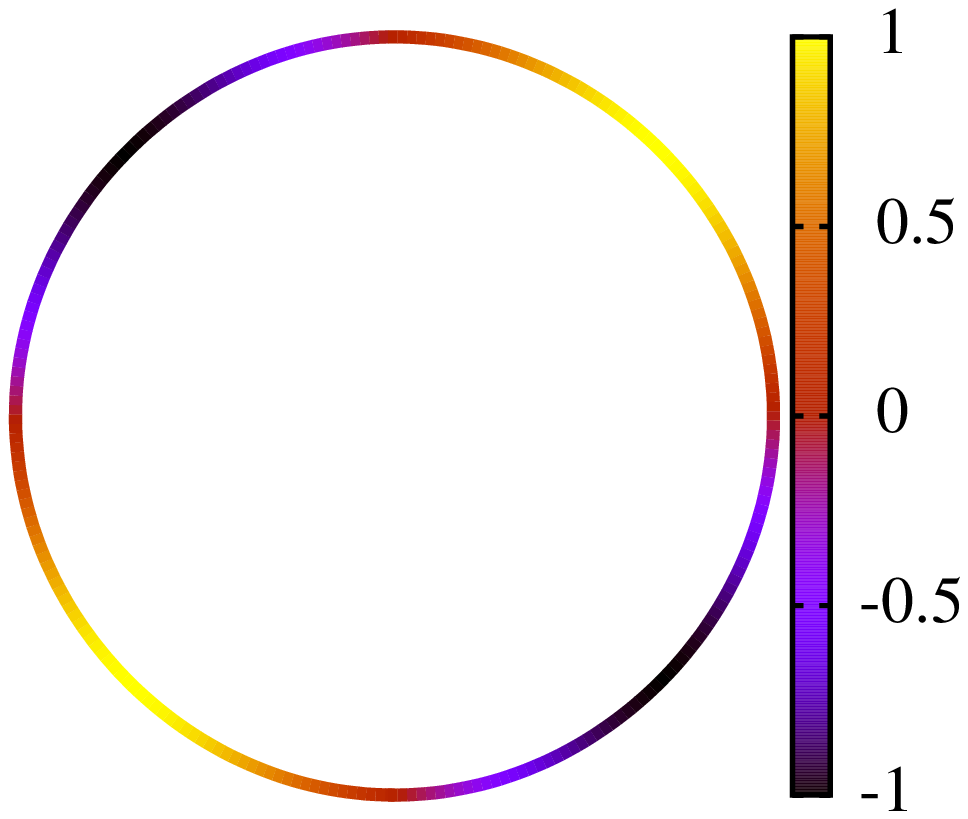}
&\centering \includegraphics[height=3.cm]{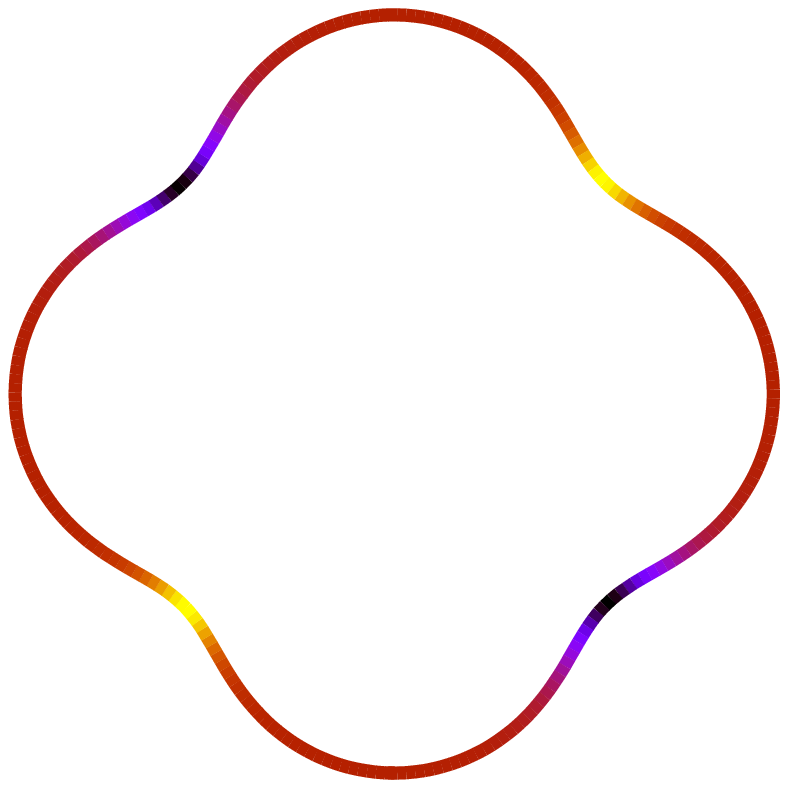}
&\centering \includegraphics[height=3.cm]{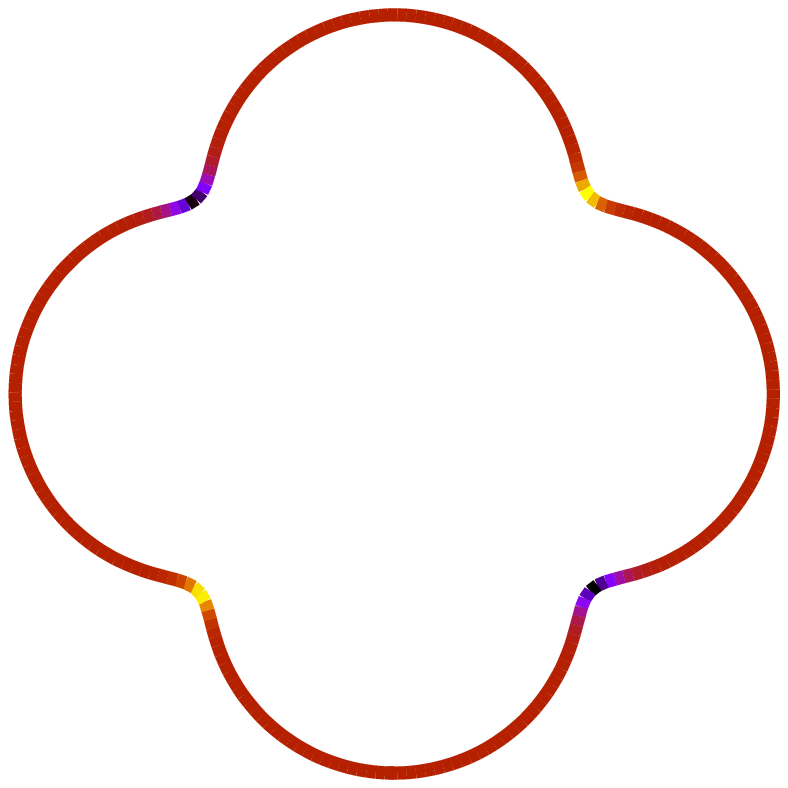}
&\\
${n_\psi=3}$& & & &\\[-4ex]
&\centering \includegraphics[height=3.cm]{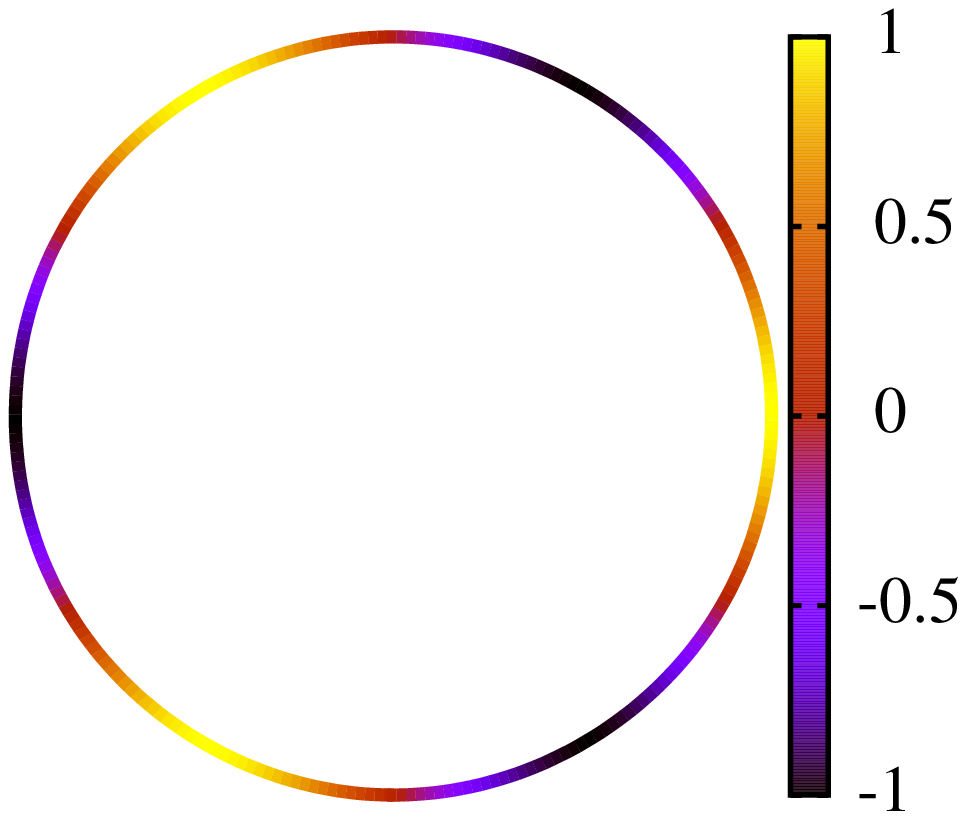}
&\centering \includegraphics[height=3.cm]{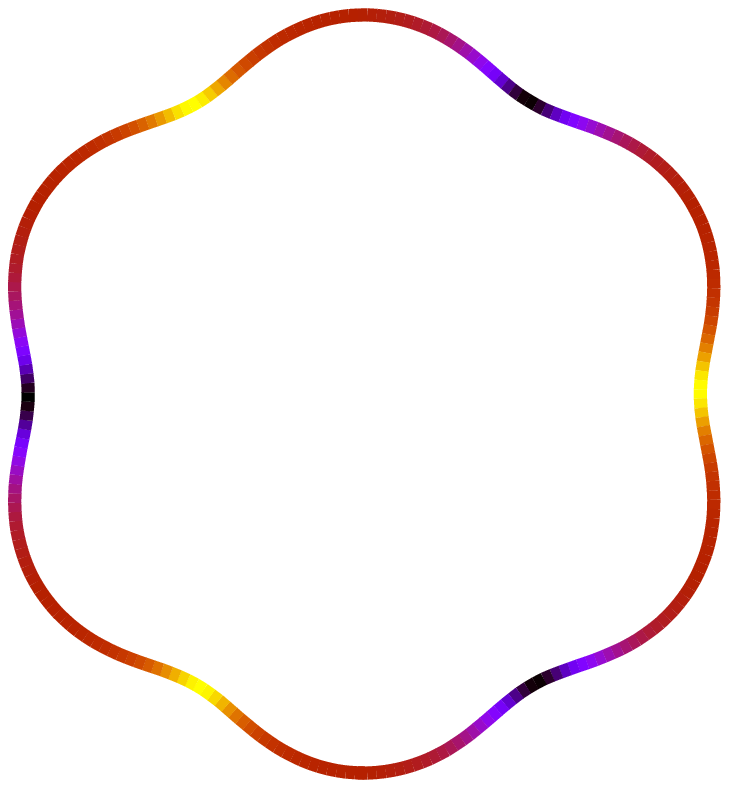}
&\centering \includegraphics[height=3.cm]{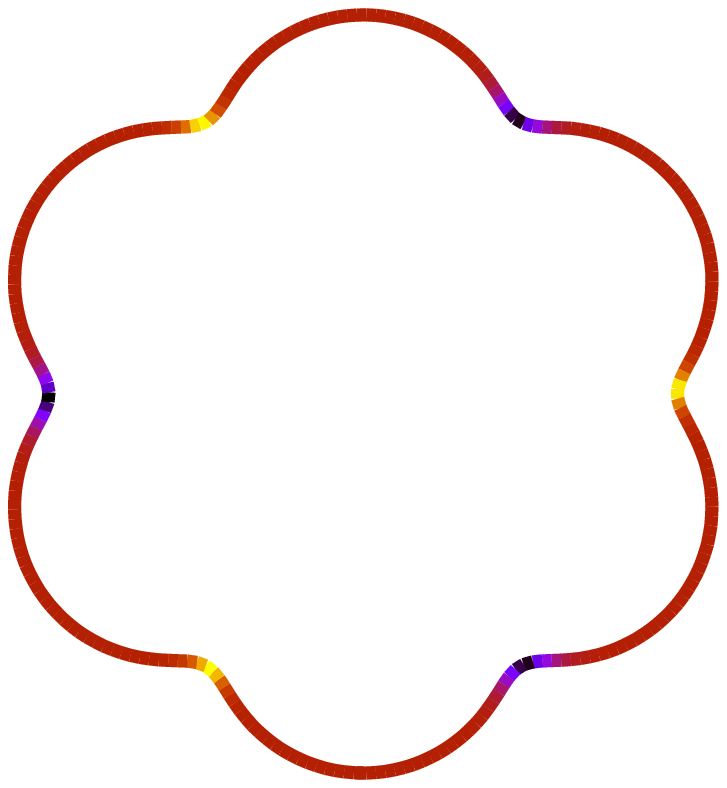}
&
\end{tabular}}
\end{table*}

First, for simplicity we consider the case when proteins are homogeneously distributed ($\rho=\const$) along the membrane and in equilibrium $\bar q=\sqrt{-2a/c}\propto\sqrt{\rho-\rho^*}$,  with $\rho>\rho^*$ (`crowded' state). The membrane contour $\gamma$ is a closed curve with constant length $L=2\pi R$, where $R\simeq 1-10~\mu$m is the characteristic size of the cell. Introducing the dimensionless quantities for length $\tilde s=s/R$ and the bending rigidity $\tilde\kappa=\kappa/(4\bar q^2\kappa_p)$, one can rewrite the sum of the energy contributions ${\cal E}_{\rm bend}+{\cal E}_{\rm LdG}+{\cal E}_{\rm anch}$ as
\be\label{eq:etot}
{\cal E}_{\rm tot}\propto \int_0^{2\pi}\!\!d\tilde s\, \bigg\{ (\p_{\tilde s} \psi+\p_{\tilde s} \theta)^2+ \tilde \kappa (\p_{\tilde s} \theta)^2 + \frac{\omega \bar q R^2\tilde \kappa}{\kappa}\sin^2(\psi-\psi_0)\bigg\}.
\ee
Note that $\p_{\tilde s}\psi$ can be thought of as spontaneous curvature, due to the coupling term $\p_{\tilde s}\psi\p_{\tilde s}\theta$, while the analogue of $\p_{\tilde s}\theta$ for surfaces is rather the Gaussian curvature than the mean curvature~\cite{docarmo:book}. In any case the form of~\eqref{eq:etot} is not equivalent to the spontaneous curvature model~\cite{helfrich:1973}. Substituting the Euler--Lagrange equation for $\theta$ we obtain the pendulum equation for $\psi$, whose solution takes the closed form
\be\label{eq:sol1}
\psi(\tilde s)=\psi_0+\am \bigg(\tilde s\, \tilde\omega\sqrt{\cal A},-\frac{1}{\cal A}\bigg),\quad
\theta(\tilde s)= {\cal B}\tilde s-\frac{\psi(\tilde s)-\psi_0}{1+\tilde\kappa},
\ee
where $\am$ is the amplitude of the Jacobi elliptic function and $\tilde\omega^2=\omega \bar q(1+\tilde\kappa) R^2/\kappa$. Two integration constants ${\cal A}$ and ${\cal B}$ are determined from the closing conditions $\psi(\tilde s+2\pi)=\psi(\tilde s)+2\pi n_\psi$ and $\theta(\tilde s+2\pi)=\theta(\tilde s)+2\pi n_\theta$, where $n_\psi,n_\theta\in \mathbb{Z}$. For nematic liquid crystals $n_\psi$ is allowed to be half-integer because of the inversion symmetry $\bp\leftrightarrow-\bp$. Here we assume that proteins bind only with a specific site to the membrane, although the collective behavior in the bulk may still be described with nematic order. Then the periodicity of $\psi(\tilde s)$ in~\eqref{eq:sol1} is $\pi_\psi=4{\cal K}(-1/{\cal A})/(\tilde\omega\sqrt{\cal A})$, where  ${\cal K}$ is a complete elliptic integral of the first kind. Then the contour length of the membrane $2\pi$ contains the integer number  $n_\psi=1,2,\ldots$ of periods $\pi_\psi$.  \looseness=-1

The resulting equilibrium shapes for a closed biomembrane are found by means of numerical integration of $x(s)=-\int_0^s d\sigma\,\cos\theta(\sigma)$ and $y(s)=\int_0^s d\sigma\,\sin\theta(\sigma)$ substituting the solution~\eqref{eq:sol1} for the angle $\theta$. The typical morphologies are presented in Table~\ref{tab:shapes}, for $\tilde\kappa=1$. The instabilities towards elongated shapes ($n_\psi=1$) and periodically undulated shapes ($n_\psi>1$) depend on a sole physical parameter $\tilde \omega$. Different colors of the closed curves measure the deviation of the actual protein tilt $\psi$ from the preferred direction $\psi_0$.  According to~\eqref{eq:sol1}, $\psi_0$ does not influence the angle $\theta$ and the corresponding shapes of curves. For relatively weak anchoring or low protein density, $\tilde\omega\simeq1$ proteins reorient together with the normal $\bn$ accordingly, and the membrane shape has a uniform curvature. For relatively strong anchoring $\tilde\omega\simeq10,30$ the proteins tend to align on average along the `easy axis' $\psi=\psi_0$ except for some narrow regions with enhanced curvature $|\p_s\theta|$, where the reorientation of $\bp$ happens.  The total energy~\eqref{eq:etot} increases with the number of buds or $n_\psi$ as listed in the Table~\ref{tab:ene}.  For a given cell size $R$, the increase of the anchoring strength $\omega_a$ or degree of protein order $\bar q$  may initiate cell elongation and the formation of buds. Thus, within the proposed framework, the anchoring forces exerted on the membrane by partly ordered proteins can drive shape instabilities.

\begin{table}[b]
\caption{\label{tab:ene}The values of ${\cal E}_{\rm tot}$ calculated with straightforward integration of~\eqref{eq:etot} for the shapes shown in Table~\ref{tab:shapes} and $\tilde\kappa=1$.}
{\def\tabcolsep{10pt}
\def\arraystretch{1.1}
\begin{tabular}{c|c cc}
${\tilde \omega}$& 1 & 10 & 30\\[1ex]
\hline
$n_\psi=1$& 20.513 & 89.425 & 249.42\\[1ex]
$n_\psi=2$& 39.294 & 160.00 & 480.00\\[1ex]
$n_\psi=3$& 70.697 & 224.461& 704.292
\end{tabular}}	
\end{table}

The presented results, based on optimizing the free energy~\eqref{eq:etot} with competing protein--protein and membrane--protein effective interactions, on a qualitative level could support  a recent protein crowding hypothesis~\cite{stachowiak:2012} as one of the plausible driving mechanisms for membrane bending. Other mechanisms include i)~membrane scaffolding by intrinsically curved proteins and ii)~insertion of amphiphatic helix into bilayer~\cite{mcmahon:2005,kozlov:2006}; both are well-studied within the spontaneous curvature models~\cite{helfrich:1973,leibler:1986,agrawal:2009,sorre:2012}. Here, however, we do not impose any direct coupling to the membrane curvature and focus on the order parameter associated with proteins. \looseness=-1

In the case of an open  membrane with small bending rigidity $\tilde \kappa\ll1 $, we can write a special solution extremizing the energy~\eqref{eq:etot}, which  is $\theta(\tilde s)\propto\arccos(\tanh(\tilde \omega\tilde s))$.  Assuming rotational symmetry along the vertical axis, this solution  resembles shape of a membrane pore (see Fig.~\ref{fig:sol}a). The color gradient marks the reorientation of proteins, which is the cause of the membrane instability. This is not rare, that solving the problem with reduced  dimensionality (1D) may capture the essential features of higher  dimensional (2D) biologically relevant shape.  However, identifying the shape instability does not mean that the proposed mechanism is the one realized in nature, where {\it in vivo} systems are essentially out-of-equilibrium. Moreover, from the modeling point of view, the assumption about homogeneous distribution of protein density is questionable. 
This leads us naturally to the subject of the following section, where we allow  spatial variation of the order parameter $q$ related to the density $\rho$ within our phenomenological {\it ansatz}.


\section{Density effects}

\begin{figure}[t]
\centering
\raisebox{20mm}{(a)}\includegraphics[width=0.95\linewidth,clip=true]{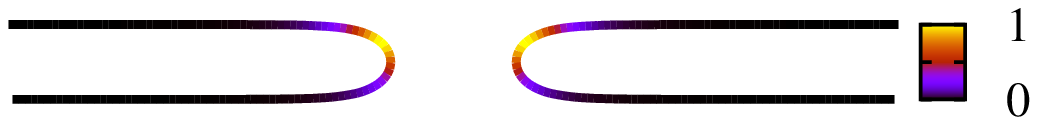}\\[-5ex]
\raisebox{25mm}{(b)}\includegraphics[width=0.95	\linewidth,clip=true]{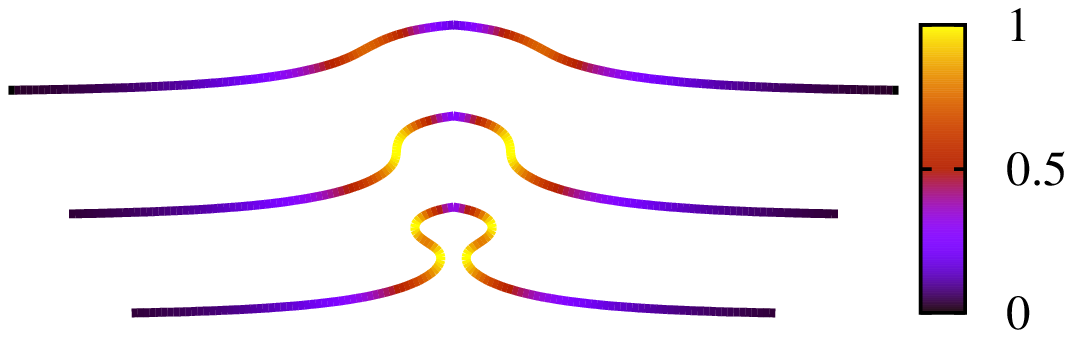}
\caption{The membrane morphologies obtained in the limit of a small bending rigidity $\tilde\kappa\ll1$. The found solutions are open curves with a mirror symmetry similar to: a)~membrane pore; b)~budding of the membrane. Membrane has a different protein coverage from top to bottom $\rho/\rho^*=(0.3,0.5,0.7)$ with maximum density at the center ($\hat q|_{s=0}=1$) and no proteins in the far field~\eqref{eq:qs}. The bending of a membrane is  computed based on the approximation of~\eqref{eq:psis}, \eqref{eq:sol2} and \eqref{eq:sol3}. The formation of bud occurs within the lengthscale $\alpha^{-1/3}$. The colorbar values show the local orientation of attached proteins in terms of $\sin(\psi-\psi_0)$.}
\label{fig:sol}
\end{figure}

Next, we aim to study the interplay between the order parameter $q$, the average orientation of proteins $\bp$ and the shape of  biomembrane~$\bn$.  Because of the non-linear coupling between these three fields, there is no hope that one can find an analytic solution to the system of partial differential equations associated with the sum of ${\cal E}_{\rm LdG}$~\eqref{eq:ldg}, ${\cal E}_{\rm bend}$ and ${\cal E}_{\rm anch}$, without assuming a certain relationships between the coefficients as well as the form of the solution. Let us study the shape instability of a flat membrane ($\theta=0$), assuming a local change of the protein order in the vicinity of $s=0$, such that $\hat q|_{s=0}=q_0>0$, while far from this point the asymptotic behavior is  $\hat q|_{s\to\infty}=\p_s\hat q|_{s\to\infty}=0$. Such perturbation can be meaningful to get additional insight into biological processes like endocytosis (budding of vesicles). The solution for $q$ can be derived from~\eqref{eq:ldg}, yielding~\cite{benamar:2011} 
\be\label{eq:qs}
\hat q(s)= \frac{2\sqrt{a/c}}{\sinh (s \sqrt{a/\kappa_p}+{\cal D})}, \quad {\cal D}=\sinh^{-1}\bigg(\frac2{q_0} \sqrt{\frac ac}\bigg),
\ee
where $\sqrt{a/c}\propto \sqrt{\rho^*-\rho}$. There exists the characteristic length scale $\sqrt{\kappa_p/a}$, where only the protein order changes, while the protein orientation $\psi$ and normal to the membrane $\theta$ assume their constant equilibrium values, which satisfy the asymptotic boundary conditions
\be
\psi|_{s\to\infty}=\psi_0,\quad  \theta|_{s\to \infty}=\p_s\theta|_{s\to \infty}=\p_s\psi|_{s\to \infty}=0. 
\ee
The effect of perturbation~\eqref{eq:qs} $q=\hat q(s)+O(\ve^2)$ can be quantified by expanding the angles $\theta$ and $\psi$ in a power series of a small parameter $\ve$, such as $\psi=\psi_0+\ve \hat\psi+O(\ve^2)$ and $\theta=\ve \hat\theta+O(\ve^2)$. In the following, we construct the approximate solution for the orientation fields $\hat\psi$ and $\hat \theta$ up to the lowest order $O(\ve)$, by linearizing the corresponding equilibrium equations, yielding
\be\label{eq:psis}
\hat \theta'=-\frac{\hat q^2\hat\psi'}{\hat q^2+\kappa/(4\kappa_p)},\quad\bigg(\frac{\hat q^2\hat \psi'}{\hat q^2+\kappa/(4\kappa_p)}\bigg)'= \frac \omega \kappa \hat q\hat \psi.
\ee
One can replace the arclength $s$ with a new variable $\sigma =\sqrt {\omega /\kappa}\int^s ds (1+\kappa/(4\kappa_p\hat q^2))$ and rewrite~\eqref{eq:psis} in conventional form
\be\label{eq:psi1}
\p_{\sigma \sigma} \hat \psi= {\cal G}[\hat q]\hat\psi, \qquad {\cal G}[\hat q]=\frac{\hat q^3}{\hat q^2 +\kappa/(4\kappa_p)}.
\ee
Then performing the integration numerically or using the WKB approximation of the one-turning point problem~\cite{bender:book} one can construct the solution to the above equation. Here, instead, we focus on the limiting case $\kappa\ll\kappa_p$, and expect the qualitative behavior of solutions to be the same. Expanding $\hat q(s)$~\eqref{eq:qs} in the neighborhood of $s=0$, we get the leading order contribution to~\eqref{eq:psis} as $\hat \psi''= \alpha (\beta-s) \hat\psi$. The solution to this equation can be expressed in terms of  the Airy function $\Ai$~\cite{bender:book} up to some constant ${\cal C}_1$
\be\label{eq:sol2}
\hat\psi_1(s)\sim{\cal C}_1\Ai\big(\alpha^{1/3}(\beta-s)\big),\quad \alpha=\frac{2q_0}{\tanh{\cal D}}\sqrt{\frac{a}{\kappa_p }}\frac \omega \kappa,
\ee
where $\beta=1/2\sqrt{a/\kappa_p}\tanh{\cal D}$ and  $\alpha^{-1/3}$ is another length scale of the problem. In the region $s\to \infty$ we have an exponentially decaying $\hat q(s)$~\eqref{eq:qs}. Then the equation~\eqref{eq:psis} reduces to $\psi''=\gamma e^{-s \sqrt{a/\kappa_p}}\psi$, whose solution can be approximated as 
\be\label{eq:sol3}
\hat\psi_2(s)\sim{\cal C}_2{\cal I}_0\bigg(2\sqrt{\frac{\gamma \kappa_p}a}e^{-s\sqrt{a/\kappa_p}}\bigg)-1,
\ee
where $\gamma=4e^{-{\cal D}}(\omega/\kappa)\sqrt{a/c}$ and ${\cal I}_0$ is the modified Bessel function of the first kind. The integration constants ${\cal C}_1$ and ${\cal C}_2$ can be found by matching two solutions~\eqref{eq:sol2} and~\eqref{eq:sol3} within a finite region or simply patching them at a single point with {\it Mathematica}~\cite{bender:book}. Thus oscillatory behavior of~\eqref{eq:sol2} is replaced with  exponentially decaying function~\eqref{eq:sol3} and the smooth solution for $\hat\psi$ is valid over the whole interval. The shape of the membrane is found by integrating numerically $\hat\theta\simeq-\hat \psi$~\eqref{eq:psis} as discussed in the previous section. The resulting curves are shown in Fig.~\ref{fig:sol}b, demonstrating the formation of bud for different protein density $\rho/\rho^*$. The budding is more pronounced at higher value of $\rho$, and the calculated shapes of curves look similar to the micrographs of membrane subjected to endocytosis~\cite{mcmahon:2005,agrawal:2009}.  As a result we found that the local inhomogeneity of the protein density~\eqref{eq:qs} may cause the  reorientation of proteins (\eqref{eq:sol2} and \eqref{eq:sol3}) and induce the shape bifurcation of biomembranes.

\section{Concluding remarks}
 
We have considered the mechanics of biomembrane coated by a layer of proteins with partial order. The presented phenomenological model captures generic morphologies of biomembrane and may contribute to understanding the origin of budding instability within the protein crowding hypothesis~\cite{stachowiak:2012}, in particular, accounting for the anchoring forces exerted on biomembrane by a layer of proteins. We believe that the proposed approach can be a step forward in establishing feasible connections between the collective behavior of proteins and geometry of membranes beyond the conventional spontaneous curvature model. Our predictions can also facilitate the development of microscopic models of membrane--protein and protein--protein interactions, which in turn would allow to determine the effective elastic constants and anchoring strength in presence of orientational order. The extension of the model to higher  dimensions (2D) as well as direct comparison between theory and experiment is the next step for the model generalization and improvement. However, projecting the protein orientation to the tangent plane of the surface and introducing coupling of the in-plane order of proteins with the geometry of a membrane is not straightforward and will require an extension of the existing models~\cite{biscari:2006,bowick:2009,napoli:2012}. 




\begin{acknowledgments}
The author is indebted to Martin Lind\'en, Jordi G\'omez-Llobregat for a number of valuable discussions and critical reading of the manuscript. The author is grateful to Aleksandr Zheltukhin for  stimulating discussions and to Dapeng~Bi for useful comments.  The author acknowledges the Soft
Matter Program at Syracuse University for financial support. 
\end{acknowledgments}

\bibliography{ref_protein}

\begin{thebibliography}{24}
\bibitem [{myn()}]{mynote1}%
  \BibitemOpen
  \href@noop {} {}\bibinfo {note} {One may still question whether this is the
  mean (extrinsic) or the Gaussian (intrinsic) curvature of biomembranes that
  proteins do sense?}\BibitemShut {Stop}%
\end{thebibliography}

\end{document}